-------------------------------------------------------------------------------

\documentstyle[12pt]{article}
\begin{document}
\baselineskip = 24pt

\begin{titlepage}
\vspace{2.5cm}
\begin{center}{\large \vspace{4.0cm} \bf The Lie algebra of the $ sl(2,{\rm
{\bf C}}) $-valued automorphic functions on a torus }\\
\vspace{1cm}
D.B.Uglov \footnotemark  \\
\footnotetext{e-mail: denis@max.physics.sunysb.edu}
\vspace{1cm}
Department of Physics, State University of New York at Stony Brook \\
Stony Brook, NY 11794-3800, USA \\
\vspace{0.5cm}
February 8 , 1993
\vspace{3cm}

\begin{abstract}
It is shown that the Lie algebra of the automorphic, meromorphic $ sl(2,{\rm
{\bf C}}) $-valued functions on a torus is a geometric realization of a certain
infinite-dimensional finitely generated Lie algebra. In the trigonometric
limit, when the modular parameter of the torus goes to zero, the former Lie
algebra goes over into the $ sl(2,{\rm {\bf C}}) $-valued loop algebra, while
the latter one - into the Lie algebra $ (A_{1}^{(1)})^{'}/(centre) $.
\end{abstract}

\end{center}
\end{titlepage}

\section{Introduction}
The Inverse Scattering Transform Method has been a source of many important
algebraic constructions. The most spectacular of them are quantum groups
associated with the quantum R-matrices, and their classical limits - the Lie
bialgebras associated with the classical r-matrices. It is well-known, that the
classical r-matrices can be classified into three categories:  rational,
trigonometric and elliptic r-matrices [2]. While the Lie bialgebras, as well as
the quantum groups, associated with the rational ( Yangians ) and the
trigonometric ( quantized Kac-Moody affine Lie algebras ) solutions of the
Yang-Baxter equation are well understood by now [4,5], not so much is known
about those associated with the elliptic R-matrices. Important exceptions are
the Lie algebras of the automorphic, meromorphic $ sl(n,{\rm {\bf C}}) $-valued
functions on a torus introduced by Reyman and Semenov-Tyan-Shanskii [1]. These
algebras are related to the elliptic solutions of the classical Yang-Baxter
equation
   , the simplest of whic
h ( $ sl(2,{\rm {\bf C}}) $-case ) is being the r-matrix of the classical
Landau-Lifshitz model [8].

The construction of Reyman and Semenov-Tyan-Shanskii is semi-geometric. Having
in view the problem of quantization it is important to give a purely algebraic
definition of their algebras in terms of finite number of generators and
defining relations. The example of what we are looking for is provided by the
trigonometric case, where the loop algebras can be considered as the geometric
realizations of the appropriate affine Kac-Moody Lie algebras [7].

As it is shown is this letter, it indeed can be done, at least in the $
sl(2,{\rm {\bf C}}) $ case. We define an infinite-dimensional, finitely
generated Lie algebra $ {\cal E}_{k,\nu^{\pm}} $ , and show that the Lie
algebra of automorphic, meromorphic $ sl(2,{\rm {\bf C}}) $-valued functons on
a torus provides a geometric realization of $ {\cal E}_{k,\nu^{\pm}} $. It will
be also shown that in the trigonometric limit, when the modular parameter of
the torus goes to zero, the Lie algebra $ {\cal E}_{k,\nu^{\pm}} $ goes over
into the Lie algebra $ {\cal L} = ( A_{1}^{(1)})^{'}/(centre) $.

It turns out that the Lie (bi)algebra $ {\cal E}_{k,\nu^{\pm}} $ can be
quantized, the corresponding quantum group being related to the eight-vertex
R-matrix.This will be the subject of the subsequent paper [9].

\section{ Definition of the Lie algebra $ {\cal E}_{k,\nu^{\pm}} $ and the main
Theorem. }

      Let $ k , k^{'} \in {\rm {\bf C}}\backslash\{0,1\} $ , and  $
k^{2}+k^{'2} = 1 $ . Let $ {\rm T} = {\rm {\bf C}}/({\rm {\bf Z}} 4K + {\rm
{\bf Z}} 4iK^{'}) $ be a torus with the periods defined by the complete
elliptic integrals $ K $ and $ K^{'} $ of the 1-st kind of the moduli $ k $ and
$ k^{'} $ correspondingly.

Introduce three meromorphic functions $ \{ w_{i} \}_{i=1,2,3} $ on $ {\rm T} $
defined in the following way [6] :

\begin{eqnarray}
w_{1}(u)=\frac{1}{sn(u)},\; w_{2}(u)=\frac{dn(u)}{sn(u)},\;
w_{3}(u)=\frac{cn(u)}{sn(u)}; \; u \in {\rm T }
\end{eqnarray}
where the Jacobi elliptic functions are of the modulus $ k $ .
These functions obey the following quadratic equations [6]:

\begin{equation}
w_{i}(u)^{2} - w_{j}(u)^{2}=J_{ij},\; i,j \in \{1,2,3\},i \neq j.
\end{equation}
where $ J_{12}=k^{2},J_{23}=k^{'2},J_{31}=-1$.

Let $ \nu^{+},\nu^{-} \in {\rm T}  $ and $  \nu^{+}-\nu^{-} \neq
n_{1}2K+n_{2}2iK^{'}  ; \; (n_{1},n_{2}) \in {\rm {\bf Z}}_{2}^{2}= {\rm {\bf
Z}/2{\bf Z} \times {\bf Z}/2{\bf Z} } $.

\begin{large}
{\bf Definition} \end{large} \hspace{1cm}
{\it
$ {\cal E}_{k,\nu^{\pm}} $ is a complex Lie algebra generated by six generators
$ \{ x_{i}^{\pm} \}_{i=1,2,3} $ which obey the following defining relations:
}

\begin{equation}
[x_{i}^{\pm},[x_{j}^{\pm},x_{k}^{\pm}]] = 0 ,
\end{equation}
\begin{equation}
[x_{i}^{\pm},[x_{i}^{\pm},x_{k}^{\pm}]]-[x_{j}^{\pm},[x_{j}^{\pm},x_{k}^{\pm}]]=J_{ij} x_{k}^{\pm},
\end{equation}
\begin{equation}
[x_{i}^+,x_{i}^{-}]=0 ,
\end{equation}
\begin{equation}
[x_{i}^{\pm},x_{j}^{\mp}] = \sqrt{-1}( w_{i}(\nu^{\mp}-\nu^{\pm})
x_{k}^{\mp}-w_{j}(\nu^{\mp}-\nu^{\pm}) x_{k}^{\pm} ) ;
\end{equation}

here, and througout the rest of the letter, $ \{ i,j,k \}$ is a cyclic
permutation  of $ \{ 1,2,3 \} $.

{\it Remark :}
The defining relations which involve both  $ x_{i}^{+} $  and  $ x_{i}^{-} $
can be compactly written in the r-matrix form, namely,
let: \[ X^{\pm}= \left( \begin{array}{cc} x_{3}^{\pm} &
x_{1}^{\pm}-ix_{2}^{\pm} \\ x_{1}^{\pm}+ix_{2}^{\pm} & -x_{3}^{\pm} \end{array}
 \right) ; \]
\begin{eqnarray}
X_{1(2)}^{\pm} = X^{\pm}\otimes I ( I \otimes X^{\pm} ) ;\;
 r_{12}(u) = \frac{1}{2} \sum_{n=1}^{3} w_{n}(u) \sigma_{n} \otimes
\sigma_{n}\: ,\: u \in {\rm T} ;
\end{eqnarray}
 where $ I $ is an identity in $ Mat_{2} $ and $ \{ \sigma_{n} \}_{n=1,2,3}$
are the Pauli matrices ; then (5,6) can be written as follows :
\begin{equation}
[X_{1}^{\pm},X_{2}^{\mp}]_{ {\cal E}_{k,\nu^{\pm}} } =
[r_{12}(\nu^{\mp}-\nu^{\pm}),X_{1}^{\pm} + X_{2}^{\mp}]_{Mat_{2}}
\end{equation}
This representation turns out to be helpful when one is doing quantization of
the Lie algebra $ {\cal E}_{k,\nu^{\pm}}  $ [9].

Now we formulate a theorem which describes the structure of the Lie algebra $
{\cal E}_{k,\nu^{\pm}}  $.
\begin{large}
{\bf Theorem}
\end{large}
\hspace{1cm} \\
1. $ {\cal E}_{k,\nu^{\pm}} = {\cal E}^{+}\oplus {\cal E}^{-} \; - \; {\cal
E}_{k,\nu^{\pm}}$ is, as a linear space, a direct sum of two Lie subalgebras
${\cal E}^{+} $ and $ {\cal E}^{-}$ generated by $ \{x^{+}_{i}\}_{i=1,2,3} $
and $ \{x^{-}_{i}\}_{i=1,2,3} $ correspondingly. \\
2. $ {\cal E}^{\pm} = \oplus_{n \in {\rm {\bf Z}}_{\geq 0} } {\cal
E}^{\pm}_{n}$ where $ dim({\cal E}^{\pm}_{n}) = 3 $.
The elements $\{g_{i}^{(n),\pm}\}_{i=1,2,3;n\in {\rm {\bf Z}}_{\geq 0}} $
defined by the recurrent formula:
\begin{equation}
g_{i}^{(0),\pm} = x_{i}^{\pm} ,
\end{equation}
\begin{equation}
g_{k}^{(n),\pm} = \frac{1}{ \sqrt{-1}n }\sum_{m+l=n-1}
[g_{i}^{(m),\pm},g_{j}^{(l),\pm}]\: , \; n\geq 1 ,\; m,l \in {\rm {\bf
Z}}_{\geq 0} , \;
\end{equation}
form a basis in $ {\cal E}^{\pm} $. The elements $
\{g_{i}^{(n),\pm}\}_{i=1,2,3}$  form a basis in $ {\cal E}^{\pm}_{n}. $  \\
3. The commutation relations of $ {\cal E}_{k,\nu^{\pm}} $ in the basis $ \{
g_{i}^{(n),\pm} \}_{i=1,2,3;n \in {\rm {\bf Z}}_{\geq 0}} $ are given by the
formulae:

{\rm a).} \begin{eqnarray} [ g_{i}^{(m),\pm}, g_{i}^{(n),\pm} ] = 0 ,& i \in
\{1,2,3 \} ,\end{eqnarray}

{\rm b).}\begin{eqnarray}
 \frac{1}{\sqrt{-1}} [ g_{i}^{(m),\pm}, g_{j}^{(n),\pm}] =
g_{k}^{(m+n+1),\pm}+(-1)^{m} \sum_{r=m}^{m+n} \left( \begin{array}{c} r \\ m
\end{array} \right) w_{r}^{i} g_{k}^{(m+n-r),\pm} - \nonumber \\ (-1)^{n}
\sum_{r=n}^{m+n} (-1)^{r} \left( \begin{array}{c} r \\ n \end{array} \right)
w_{r}^{j} g_{k}^{(m+n-r),\pm} ,
\end{eqnarray}

{\rm c).} \begin{eqnarray} [ g_{i}^{(m),+}, g_{i}^{(n),-} ] = 0 ,& i \in
\{1,2,3 \},\end{eqnarray}

{\rm d).}\begin{eqnarray}
 \frac{1}{\sqrt{-1}} [ g_{i}^{(m),\pm}, g_{j}^{(n),\mp}] = (-1)^{m}
\sum_{r=m}^{m+n} \left( \begin{array}{c} r \\ m \end{array} \right)
v_{r}^{i}(\pm (\nu^{-}-\nu^{+})) g_{k}^{(m+n-r),\mp} - \nonumber \\ (-1)^{n}
\sum_{r=n}^{m+n} (-1)^{r} \left( \begin{array}{c} r \\ n \end{array} \right)
v_{r}^{j}(\pm(\nu^{-}-\nu^{+})) g_{k}^{(m+n-r),\pm} ,
\end{eqnarray}
$ m,n \in {\rm {\bf Z}}_{\geq 0} $, and the coefficients $ w_{r}^{i} ,
v_{r}^{i}(\nu) $ are defined by the formulae:
\begin{eqnarray}
w_{i}(u)= \frac{1}{u}+\sum_{r \geq 0}w_{r}^{i} u^{r}\;, \;
v_{r}^{i}(u)= \frac{1}{n!} \frac{d^{n}}{du^{n}} w_{i}(\nu).
\end{eqnarray}

Proof of the theorem is given in the Appendix.

{\it Remark :} Commutation relations in $ {\cal E}_{k,\nu^{\pm}} $ can be
compactly written down using the generating functions and the r-matrix
representation. Let us define:
\begin{equation}
G_{i}^{\pm}(\alpha)=\sum_{n \geq 0} \alpha^{n} g_{i}^{(n),\pm} \; , \:
\alpha\in {\rm T};\end{equation}
\begin{equation}
{\bf G^{\pm}} = \left( \begin{array}{cc} G_{3}^{\pm} & G_{1}^{\pm}-iG_{2}^{\pm}
\\ G_{1}^{\pm}+iG_{2}^{\pm} & -G_{3}^{\pm} \end{array}  \right) ;
\end{equation}
then the commutation relations in the basis $ \{ g_{i}^{(n),\pm} \} $ can be
recovered from the following commutation relations for the generating
functions:
\begin{equation}
[{\bf G}_{1}^{\pm}(\alpha),{\bf G}_{2}^{\pm}(\beta)]_{{\cal E}_{k,\nu^{\pm}}} =
[r_{12}(\beta-\alpha),{\bf G}_{1}^{\pm}(\alpha)+{\bf
G}_{2}^{\pm}(\beta)]_{Mat_{2}}. \end{equation}
\begin{equation}
[{\bf G}_{1}^{\pm}(\alpha),{\bf G}_{2}^{\mp}(\beta)]_{{\cal E}_{k,\nu^{\pm}}} =
[ r_{12}(\beta-\alpha+(\nu^{\mp}-\nu^{\pm})),{\bf G}_{1}^{\pm}(\alpha)+{\bf
G}_{2}^{\mp}(\beta)]_{Mat_{2}}. \end{equation}

\section{Geometric realization of $ {\cal E}_{k,\nu^{\pm}}  $ }

In this section we describe a geometric realization of the Lie algebra $ {\cal
E}_{k,\nu^{\pm}}  $ - an ``elliptic analog'' of the loop Lie algebras. This
turns out to be the Lie algebra of the  $ sl(2,{\rm {\bf C}}) $-valued
automorphic meromorphic functions on T  [1].

Consider a set $ \tilde{F} =  \{ \varphi_{i}^{(n),\pm} \}_{i=1,2,3},n \in {\rm
{\bf Z}}_{\geq 0} $ of meromorphic functions on T defined as follows:
\begin{equation}
\varphi_{i}^{(n),\pm}(u) = \frac{(-1)^{n}}{n!} \frac{d^{n}}{du^{n}}
w_{i}(u-\nu^{\pm}), \; u \in {\rm T}
\end{equation}

Let ${\cal G}=sl(2,{\rm {\bf C}})$. Fix a basis $ \{ s_{i} \}_{i=1,2,3} $ in
${\cal G} $ such that: $ [ s_{i},s_{j} ] = \sqrt{-1}s_{k} $. Let {\it F} be a
linear span of $ \tilde{F} $ over {\bf C} . Consider a tensor product $ {\cal
A}={\it F} \otimes_{\rm {\bf C}} {\cal G} $. Single out a linear subspace of $
{\cal A} $  spanned by the elements of the form $ \varphi_{i}^{(n),\pm} \otimes
s_{i} $. Denote this subspace by $ \tilde{{\cal E}_{k,\nu^{\pm}}} $. The
elements of this subspace satisfy the following automorphicity condition with
respect to the action of the group $ {\rm {\bf Z}^{2}_{2}} = {\rm {\bf Z}/2{\bf
Z} \times {\bf Z}/2{\bf Z} } $:
\begin{equation}
\varphi_{i}^{(n),\pm}(u+n_{1}2K+n_{2}i2K^{'}) \otimes s_{i} =
\varphi_{i}^{(n),\pm}(u) \otimes
(T^{(n_{1},n_{2})}s_{i}(T^{(n_{1},n_{2})})^{-1})\;,\; (n_{1},n_{2}) \in {\rm
{\bf Z}_{2}^{2}}   \end{equation}
where, in the $ 2 \times 2 $ matrix realization of $ {\cal G} $ , one has $
T^{(n_{1},n_{2})}=\sigma_{3}^{n_{1}} \sigma_{1}^{n_{2}} $ .
 Introduce a Lie bracket in  $ \tilde{{\cal E}_{k,\nu^{\pm}}} $ as follows: $ [
\varphi_{i}^{(n),\epsilon} \otimes s_{i}, \varphi_{j}^{(n),\epsilon^{'}}
\otimes s_{j}]= \varphi_{i}^{(n),\epsilon} \varphi_{j}^{(n),\epsilon^{'}}
\otimes [s_{i},s_{j}];\; i,j=1,2,3.;\: \epsilon,\epsilon^{'} \in \{+,-\} $.
Then the map $ \gamma :  {\cal E}_{k,\nu^{\pm}} \rightarrow  \tilde{{\cal
E}_{k,\nu^{\pm}}} \;  :  \;  \gamma(g_{i}^{(n),\pm}) =
\varphi_{i}^{(n),\pm}\otimes s_{i} $ is an isomorphism, as can be readily
checked using the relations :
\begin{equation}
 w_{i}(u)w_{j}(v)=w_{k}(u-v)w_{i}(v)-w_{k}(u-v)w_{j}(u) \; , \; u,v \in {\rm
T}\; {\rm [6]} .
\end{equation}

\newtheorem{prop}{Proposition}

\begin{prop}
$ {\cal E}_{k,\nu^{\pm}} $ is a Lie bialgebra { \rm [3,4] } , cocommutator map
$ \delta : {\cal E}_{k,\nu^{\pm}} \rightarrow \bigwedge^{2} {\cal
E}_{k,\nu^{\pm}} $ is being given by the following formulae :
\begin{equation}
\delta[ g_{i}^{(n),\pm} ] = \sqrt{-1} \sum_{l=0}^{n} g_{k}^{(n-l),\pm}\bigwedge
g_{j}^{(l),\pm},
\end{equation}
 or, in terms of the generating functions:
\begin{equation}
\delta[ G_{i}^{\pm}(\cdot,\alpha)](u,v)=[G_{i}^{\pm}(u,\alpha)\otimes I +
I\otimes G_{i}^{\pm}(v,\alpha),r(u-v)]_{U{\cal G}\otimes U{\cal G}}
\end{equation}

where $ G_{i}^{\pm}(u,\alpha) = \gamma(G_{i}^{\pm}(\alpha))=
w_{i}(u-\alpha-\nu^{\pm}) \otimes s_{i} ,
r(u)= 2 \sum_{k=1}^{3} w_{k}(u) s_{k}\otimes s_{k} \; \in U{\cal G}\otimes
U{\cal G}, \: u\in {\rm T} $ , and $ I $ is an identity in $ U{\cal G} $-the
universal enveloping algebra of $ {\cal G} $.

\end{prop}

\section{Trigonometric limit of $ {\cal E}_{k,\nu^{\pm}} $ }

In this section it is shown that the trigonometric limit: $ k \rightarrow 0 $
of $ {\cal E}_{k,\nu^{\pm}} $, when $ \nu^{+} = i\frac{3}{2}K^{'}
\;,\nu^{-}=i\frac{1}{2}K^{'} $ coincides with the $ sl(2) $-loop algebra : $
{\cal L}=(\hat{{\cal G}})^{'}/(centre)$ , where $ \hat{{\cal G}}=A_{1}^{(1)}$.

Let us fix $ \nu^{+} = i\frac{3}{2}K^{'} \;,\nu^{-}=i\frac{1}{2}K^{'} $ , let
us also make a change of the generators of $ {\cal E}_{k,\nu^{\pm}} $ :
$ x_{i}^{\pm}\rightarrow x_{i}^{'\pm} $ , where $
x_{1,2}^{'\pm}=\frac{1}{\sqrt{k}}x_{1,2}^{\pm} ;\:x_{3}^{'\pm}=x_{3}^{\pm}.$
Then the defining relations of $ {\cal E}_{k,\nu^{\pm}} $ acquire a form (from
now on we drop the prime):
\begin{eqnarray}
 [x_{1}^{+},x_{2}^{-}]=x_{3}^{+},& [x_{1}^{-},x_{2}^{+}]=-x_{3}^{-},&
[x_{2}^{+},x_{3}^{-}]=x_{1}^{+}-k x_{1}^{-},
\end{eqnarray}
\begin{eqnarray}
[x_{2}^{-},x_{3}^{+}]=-x_{1}^{-}+k x_{1}^{+},& [x_{3}^{+},x_{1}^{-}]=-
x_{2}^{-},& [x_{3}^{-},x_{1}^{+}]= x_{2}^{+};
\end{eqnarray}
\begin{eqnarray}
[x_{1}^{\pm},[x_{1}^{\pm},x_{3}^{\pm}]]-[x_{2}^{\pm}
,[x_{2}^{\pm},x_{3}^{\pm}]]=k x_{3}^{\pm}, & k
[x_{2}^{\pm},[x_{2}^{\pm},x_{1}^{\pm}]]-[x_{3}^{\pm},[x_{3}^{\pm},x_{1}^{\pm}]]=k^{'2} x_{1}^{\pm} ,
\end{eqnarray}
\begin{eqnarray}
[x_{3}^{\pm},[x_{3}^{\pm},x_{2}^{\pm}]]-
k[x_{1}^{\pm},[x_{1}^{\pm},x_{2}^{\pm}]]=-x_{2}^{\pm} ,&
[x_{i}^{\pm},[x_{j}^{\pm},x_{k}^{\pm}]]=0,\; [x_{i}^{+},x_{i}^{-}]=0,\: i=1,2,3
{}.
\end{eqnarray}

If $ k\neq 0 $ , it follows from the above relations that
\begin{eqnarray}
[x_{2}^{\pm},[x_{1}^{\pm},[x_{1}^{\pm},x_{2}^{\pm}]]]=0,& {\rm and} \; \;
[x_{2}^{\pm},[x_{2}^{\pm},[x_{2}^{\pm},x_{1}^{\pm}]]]+[x_{1}^{\pm},[x_{1}^{\pm},[x_{1}^{\pm},x_{2}^{\pm}]]]= k [x_{1}^{\pm},x_{2}^{\pm}]
\end{eqnarray}

Let us add these new ( non-independent only if $ k\neq 0 $) relations to the
relations (25-28) and then take a limit $ k\rightarrow 0 $. As a result $ {\cal
E}_{k,\nu^{\pm}} $ goes over into the Lie algebra $ \tilde{{\cal L}}$ generated
by the elements $ \{y_{i}^{\pm}=lim_{k\rightarrow 0}x_{i}^{\pm} \}_{i=1,2,3} $.
In the geometric realization this limit has the following explicit form:
\begin{eqnarray}
\frac{1}{\sqrt{k}}w_{1}(u-\nu^{\pm})\otimes s_{1}=\sqrt{k}sn(u \mp
\frac{i}{2}K^{'}) \otimes s_{1} \rightarrow \mp ie^{\pm iu} \otimes s_{1} ,
\nonumber \\
\frac{1}{\sqrt{k}}w_{2}(u-\nu^{\pm})\otimes s_{2}=i\sqrt{k}cn(u \mp
\frac{i}{2}K^{'}) \otimes s_{2} \rightarrow ie^{\pm iu} \otimes s_{2} ,
\nonumber \\
w_{3}(u-\nu^{\pm})\otimes s_{3}=idn(u \mp \frac{i}{2}K^{'}) \otimes s_{3}
\rightarrow i\otimes s_{3} . \nonumber
\end{eqnarray}
$ \{y_{i}^{\pm} \} $ obey the following defining relations:
\begin{eqnarray}
[y_{1}^{\pm},y_{2}^{\mp}]=\pm y_{3}^{\pm},& [y_{2}^{\pm},y_{3}^{\mp}]=\pm
y_{1}^{\pm},& [y_{3}^{\pm},y_{1}^{\mp}]=\mp y_{2}^{\mp},
\end{eqnarray}
\begin{eqnarray}
[y_{1}^{\pm},[y_{1}^{\pm},y_{3}^{\pm}]]-[y_{2}^{\pm},[y_{2}^{\pm},y_{3}^{\pm}]]=0 ,& [y_{3}^{\pm},[y_{3}^{\pm},y_{2}^{\pm}]]=-y_{2}^{\pm},& [y_{3}^{\pm},[y_{3}^{\pm},y_{1}^{\pm}]]=-y_{1}^{\pm},
\end{eqnarray}
\begin{eqnarray}
[y_{i}^{\pm},[y_{j}^{\pm},y_{k}^{\pm}]]=0,& [y_{i}^{+},y_{i}^{-}]=0,&
[y_{2}^{\pm},[y_{1}^{\pm},[y_{1}^{\pm},y_{2}^{\pm}]]]=0 ,
\end{eqnarray}
\begin{eqnarray}
[y_{2}^{\pm},[y_{2}^{\pm},[y_{2}^{\pm},y_{1}^{\pm}]]]+[y_{1}^{\pm},[y_{1}^{\pm},[y_{1}^{\pm},y_{2}^{\pm}]]]=0.
\end{eqnarray}

\begin{prop}
$ \tilde{{\cal L}} $ contains an ideal $\imath$ generated by the elements:
\begin{eqnarray}
[y_{2}^{\pm},y_{3}^{\pm}]\mp y_{1}^{\pm},&[y_{3}^{\pm},y_{1}^{\pm}]\mp
y_{2}^{\pm},& y_{3}^{+}-y_{3}^{-}
\end{eqnarray}
\end{prop}

\begin{prop}
Lie algebra $ {\cal L }=\tilde{{\cal L}}/\imath $ is isomorphic to the $
sl(2,{\rm {\bf C}})$-loop Lie algebra $ (\hat{{\cal G}})^{'}/(centre)$ ,
$\hat{{\cal G}}=A_{1}^{(1)}$.
\end{prop}
{\it Remark:} Non-independent relations (29) in $ {\cal E}_{k,\nu^{\pm}} $ go
over , as $ k\rightarrow 0 $ into the Serre relations in $ {\cal L}$.

\section{Conclusion}
The fact, that the Lie algebra of the automorphic, meromorphic $ sl(2,{\rm {\bf
C}}) $-valued functions on a torus admits a purely algebraic description as the
finitely generated infinite dimensional Lie algebra, suggests that it is
possible to develop a theory of the algebras of such a type which would be
parallel to the theory of the Kac-Moody affine Lie algebras [7]. Such a
developement would be clearly important taking into account an abundance of
physical and mathematical constructions related to the latter. Among the more
immediate ramifications of the result given in this letter, one can point out
the problem of finding the quantum groups related to the elliptic quantum
R-matrices [9].

{\bf Acknowlegement} \\
The author is grateful to I.T.Ivanov and L.A.Takhtadjan for many helpful
discussions. \\
\vspace{1cm}

{\bf Appendix}

{\it  Proof of the theorem .} From the defining relations (3-6) and the Jacobi
identity it follows that $ {\cal E}_{k,\nu^{\pm}} = {\cal E}^{+} \cup {\cal
E}^{-} $, where ${\cal E}^{+}$ and $ {\cal E}^{-} $ are Lie subalgebras
generated by $ \{ x_{i}^{+} \}_{i=1,2,3} $ and $ \{ x_{i}^{-} \}_{i=1,2,3}  $
correspondingly.

Consider $ {\cal E}^{+} $ . $ {\cal E}^{+} = \cup_{n \geq 0} \overline{{\cal
E}}_{n}^{+} $, where $ \overline{{\cal E}}_{n}^{+} $ is a linear subspace of $
{\cal E}^{+} $ spanned by all the elements of the form $ [ x_{i_{1}}^{+},
\ldots , [ x_{i_{n}}^{+} , x_{i_{n+1}}^{+} ] \ldots ],\; i_{p} \in \{ 1,2,3 \}
$ if $ n \geq 1 $, and $ \overline{{\cal E}}_{0}^{+} $ is a linear span of $ \{
x_{i}^{+} \}_{i=1,2,3} $.

In what follows $ g_{i}^{(n)} \stackrel{\rm def}{=} g_{i}^{(n),+}, \; i \in
\{1,2,3 \}, \; n \geq 0.$ We remind, that the indices $ \{ i,j,k \} $ denote
any cyclic permutation of $ \{1,2,3 \}$. We will prove the statements 2 and 3
a).,b). of the theorem using induction.

Fix $ n \in {\rm {\bf Z}},\; n \geq 2 $ and assume, that:

$ {\rm A}_{n} $.  The elements $ \{g_{i}^{(s)} \}_{i=1,2,3},\;0 \leq s \leq n $
form a basis in $ \cup_{s=0}^{n}\overline{{\cal E}}_{s}^{+} $. Denote by $
{\cal E }_{s}^{+} $ a linear subspace of $ \cup_{s=0}^{n}\overline{{\cal
E}}_{s}^{+} $ spanned by the elements $ \{g_{i}^{(s)} \}_{i=1,2,3} $. We have $
\cup_{m=0}^{n} \overline{{\cal E}}_{m}^{+} = \oplus_{m=0}^{n} {\cal E}_{m}^{+}
$.

$ {\rm B}_{n} $.  $ [g_{i}^{(l)}, g_{i}^{(m)}] = 0 ,\; $ if $ \; 0 \leq l+m
\leq n-1 $.

$ {\rm C}_{n} $.  There exist the coefficients $ A_{k,r}^{(l,m)} $ such, that
\begin{equation}
\frac{1}{\sqrt{-1}} [g_{i}^{(l)},g_{j}^{(m)}] = g_{k}^{(l+m+1)} +
\sum_{r=0}^{l+m-1} A_{k,r}^{(l,m)} g_{k}^{(r)}, \end{equation} if $ 0 \leq l+m
\leq n-1 $ .

The fact that these assumptions are true for $ n=2 $ , and that $
A_{k,0}^{(0,1)} = \frac{1}{2}J_{ij} = w_{1}^{i}-w_{1}^{j} \; , \;
A_{k,0}^{(1,0)} = -\frac{1}{2}J_{ij} $, follows immediately from the defining
relations (3,4). We will show that under the above assumptions the statements $
{\rm A}_{n+1},\;{\rm B}_{n+1},\; {\rm C}_{n+1} $ hold true .

{}From $ {\rm A}_{n} $ it follows that $ \overline{{\cal E}}^{+}_{n+1} $ is a
linear span of $ \{ [g_{i}^{(m)},g_{i^{'}}^{(n-m)}] \}_{i,i^{'} \in \{ 1,2,3
\}; 0\leq m \leq n}.$ Let us show that among the elements $
\{[g_{i}^{(m)},g_{j}^{(n-m)}] \}_{0\leq m \leq n} $ only three are linearly
independent mod($ \oplus_{s=0}^{n} {\cal E}_{s}^{+} $). This fact follows from
the:

{\bf Lemma 1.} Under the assumptions $ {\rm A}_{n},{\rm B}_{n},{\rm C}_{n} $:

1. $ [g_{i}^{(m+1)},g_{j}^{(n-m-1)}] = [g_{i}^{(m)},g_{j}^{(n-m)}] +
\eta_{k}^{(n,m)} \; , \; 0 \leq m \leq n-1 ,$ where \begin{equation}
\eta_{k}^{(n,m)} = \sqrt{-1} \sum_{s=0}^{n-1} B_{k,s}^{(n,m)} g_{k}^{(s)} \;
\in \; \oplus_{r=0}^{n-1} {\cal E}_{r}^{+}. \end{equation}

2. The coefficients $ B_{k,s}^{(n,m)} $ are expressed in terms of the
coefficients $ \{ A_{k,s}^{(l,m)} \}_{0 \leq l+m \leq n-1} $ by the formulae:
\begin{eqnarray}
B_{k,s}^{(n,m)} = \theta (s-m-1) A_{j,s-m-1}^{(n-2m-2,m+1)} - \theta (s-m-2)
A_{j,s-m-2}^{(n-2m-2,m)} + \nonumber \\ \sum_{r=0}^{n-m-2} \theta (m+r-1-s)
A_{j,r}^{(n-2m-2,m+1)} A_{k,s}^{(m,r)} - \sum_{r=0}^{n-m-3} \theta (m+r-s)
A_{j,r}^{(n-2m-2,m)} A_{k,s}^{(m+1,r)} ,\; \nonumber \\ if\; 0 \leq m \leq p-1
, \; n=2p \; or \; n=2p+1;
\end{eqnarray}
\begin{eqnarray}
B_{k,s}^{(n,m)} =  \theta (s-n+m-1) A_{i,s-n+m-1}^{(n-m-1,2m-n)} - \theta
(s-n+m) A_{i,s-n+m}^{(n-m,2m-n)} +\nonumber \\
\sum_{r=0}^{m-2} \theta (r+n-m-1-s) A_{i,r}^{(n-m-1,2m-n)} A_{k,s}^{(r,n-m)} -
\nonumber \\ \sum_{r=0}^{m-1} \theta (r+n-m-2-s) A_{i,r}^{(n-m,2m-n)}
A_{k,s}^{(r,n-m-1)} , \nonumber \\ if\; p \leq m \leq n=2p , \; or \; p+1 \leq
m \leq n=2p+1;
\end{eqnarray}
\begin{eqnarray}
B_{k,s}^{(2p+1,p)} = \theta (s-p-2) A_{i,s-p-2}^{(p-1,0)} - \theta (s-p)
A_{i,s-p}^{(p+1,0)} + \sum_{r=0}^{p-2} \theta (r+p-s) A_{i,r}^{(p-1,0)}
A_{k,s}^{(r,p+1)} - \nonumber \\ \sum_{r=0}^{p} \theta (r+p-2-s)
A_{i,r}^{(p+1,0)} A_{k,s}^{(r,p-1)} - B_{k,s}^{(2p+1,p+1)},
where \; \theta (x \geq (<) 0) = 1(0).
\end{eqnarray}

{\it Proof}. Let $ n=2p $ or $ n=2p+1 $.

Let  $ 0 \leq m \leq p-1 $ . According to the assumption $ {\rm C}_{n} $:
\begin{eqnarray}
[g_{i}^{(m)},g_{j}^{(n-m)}] = \sqrt{-1}
[g_{i}^{(m)},[g_{i}^{(m+1)},g_{k}^{(n-2m-2)}]] + \vartheta_{k}^{(n,m)},\\
-\vartheta_{k}^{(n,m)} = \sqrt{-1}(\sum_{r=0}^{n-m-1} A_{j,r}^{(n-2m-2,m+1)}
g_{k}^{(m+r+1)}+\sum_{r=0}^{n-m-2} A_{j,r}^{(n-2m-2,m+1)} \sum_{s=0}^{m+r-1}
A_{k,s}^{(m,r)} g_{k}^{(s)}).
\end{eqnarray}
It is clear, that $ \vartheta_{k}^{(n,m)} \in \oplus_{r=0}^{n-1} {\cal
E}_{r}^{+} $.
\begin{eqnarray}
[g_{i}^{(m+1)},g_{j}^{(n-m-1)}] = \sqrt{-1}
[g_{i}^{(m+1)},[g_{i}^{(m)},g_{k}^{(n-2m-2)}]] + \vartheta_{k}^{(n,m+1)},\\
-\vartheta_{k}^{(n,m+1)} = \sqrt{-1}( \sum_{r=0}^{n-m-3} A_{j,r}^{(n-2m-2,m)}
g_{k}^{(m+r+2)}+ \sum_{r=0}^{n-m-3} A_{j,r}^{(n-2m-2,m)} \sum_{s=0}^{m+r}
A_{k,s}^{(m+1,r)} g_{k}^{(s)})\: \in  \oplus_{r=0}^{n-1} {\cal
E}_{r}^{+}.\end{eqnarray}
Using the Jacobi identity and the  assumption $ {\rm B}_{n} $ we get
\begin{equation} [g_{i}^{(m)},[g_{i}^{(m+1)},g_{j}^{(n-2m-2)}]] =
[g_{i}^{(m+1)},[g_{i}^{(m)},g_{j}^{(n-2m-2)}]].\end{equation}
 Thus $ [g_{i}^{(m+1)},g_{j}^{(n-m-1)}] = [g_{i}^{(m)},g_{j}^{(n-m)}] +
\eta_{k}^{(n,m)}, \; \eta_{k}^{(n,m)} = \vartheta _{k}^{(n,m+1)} -  \vartheta
_{k}^{(n,m)}\;  \in  \oplus_{r=0}^{n-1} {\cal E}_{r}^{+} $.

The statement 1. is proven in the case $ 0 \leq m \leq p-1 $ . The statement 2.
in this case follows from the above formulae for $ \eta_{k}^{(n,m)} $. The
cases $ p \leq m \leq n=2p \; ; p+1 \leq m \leq n=2p+1 $ and $ m=p ,\; n=2p+1 $
are proven analogously  $\Box $.

{}From the Lemma 1. it follows that $ [g_{i}^{(m)},g_{j}^{(n-m)}] =
[g_{i}^{(0)},g_{j}^{(n)}] + \sum_{r=0}^{m-1} \eta_{k}^{(n,r)} ,\; 1 \leq m \leq
n $. By the definition of $ g_{k}^{(n+1)} $ we also have: $ \sqrt{-1} (n+1)
g_{k}^{(n+1)} = \sum_{m=0}^{n} [g_{i}^{(m)},g_{j}^{(n-m)}] $. Hence we obtain
the following expressions for the commutators $ [g_{i}^{(m)},g_{j}^{(n-m)}] $:
\begin{equation}
[g_{i}^{(0)},g_{j}^{(n)}] = \sqrt{-1} g_{k}^{(n+1)} - \frac{1}{n+1}
\sum_{s=1}^{n}\sum_{r=0}^{s-1} \eta_{k}^{(n,r)} \; ,
\end{equation}
\begin{eqnarray}
[g_{i}^{(m)},g_{j}^{(n-m)}] = \sqrt{-1} g_{k}^{(n+1)} + \sum_{r=0}^{m-1}
\eta_{k}^{(n,r)} - \frac{1}{n+1} \sum_{s=1}^{n}\sum_{r=0}^{s-1}
\eta_{k}^{(n,r)} \; ,\; 1 \leq m \leq n .
\end{eqnarray}
{}From these expressions and the fact, that $ [g_{i}^{(m)},g_{i}^{(n-m)}] = 0
\;,
\; 0 \leq m \leq n ; \; i \in \{ 1,2,3 \} $, which will be proven in the Lemma
2., it follows, that $ \{ g_{i}^{(n+1)} \}_{i=1,2,3} $ form a basis in $
\overline{{\cal E}}_{n+1}^{+}/ \cup_{0 \leq r \leq n} \overline{{\cal
E}}_{r}^{+} $ . Thus we have proven that $ {\rm A}_{n+1} $ holds. From (45,46)
and the second statement of the Lemma 1. it also follows, that $ {\rm C}_{n+1}
$ is true . The coefficients $ A_{k,r}^{(m,n-m)} $ are expressed in terms of
the coefficients $ \{ A_{k,r}^{(l,m)} \}_{0 \leq l+m \leq n-1 } $ by the
following formulae:
\begin{eqnarray}
A_{k,s}^{(m,n-m)} = (1-\delta_{0m}) \sum_{r=0}^{m-1} B_{k,s}^{(n,r)} -
\frac{1}{n+1} \sum_{r=0}^{n-1}(n-r)B_{k,s}^{(n,r)} \; , \; 0 \leq m \leq n \; ,
\; 0 \leq s \leq n-1 .
\end{eqnarray}
Recall, that the coefficients $ B_{k,s}^{(n,r)} $ are expressed in terms of $
\{ A_{i,s}^{(l,m)} \}_{i=1,2,3; 0 \leq l+m \leq n-1 } $ as given by the second
statement of the Lemma 1. Combining the formulae (37-39) and (47) we obtain
recurrent relations for $ A_{k,s}^{(l,m)} $.From these relations it is evident,
that all the coefficients  $ A_{k,s}^{(l,m)} $ are completely and uniquely
determined by $ A_{k,0}^{(0,1)} $ and $ A_{k,0}^{(1,0)} $ which enter into the
defining relations (3,4).

Let us now prove $ {\rm B}_{n+1} $.

{\bf Lemma 2.} Under the assumptions $ {\rm A}_{n},{\rm B}_{n},{\rm C}_{n} $:

$ [g_{i}^{(m)},g_{i}^{(n-m)}] = 0 \; ; \; i \in \{ 1,2,3 \}, \; 0 \leq m \leq n
. $

{\it Proof.} First, let us prove, that $ [g_{i}^{(0)},g_{i}^{(n)}] = 0 $.
{}From the assumptions $ {\rm B}_{n} $  and  $ {\rm C}_{n} $  we have:
\begin{eqnarray}
\sqrt{-1}[g_{i}^{(0)},g_{i}^{(n)}] = [g_{i}^{(0)},[g_{j}^{(0)},g_{k}^{(n-1)}]]
=  [g_{i}^{(0)},[g_{j}^{(1)},g_{k}^{(n-2)}]] = \ldots =
[g_{i}^{(0)},[g_{j}^{(n-1)},g_{k}^{(0)}]] .
\end{eqnarray}
By the Jacobi identity and the assumptions $ {\rm B}_{n} $ , $ {\rm C}_{n} $
we also have:
\begin{eqnarray}
[g_{i}^{(0)},g_{i}^{(n)}] = - [g_{j}^{(m)},g_{j}^{(n-m)}] -
[g_{k}^{(n-m-1)},g_{k}^{(m+1)}].
\end{eqnarray}
{}From (48) and (49) we get:
\begin{eqnarray}
n [g_{i}^{(0)},g_{i}^{(n)}] = - \sum_{m=0}^{n-1}([g_{j}^{(m)},g_{j}^{(n-m)}] +
[g_{k}^{(n-m-1)},g_{k}^{(m+1)}]) = -  [g_{j}^{(0)},g_{j}^{(n)}] -
[g_{k}^{(0)},g_{k}^{(n)}] ,
\end{eqnarray}
and
\begin{equation}
(n-1) ( [g_{i}^{(0)},g_{i}^{(n)}] - [g_{j}^{(0)},g_{j}^{(n)}] ) = 0 .
\end{equation}
Recall, that we  consider only $  n \geq 2 $, so : $ [g_{i}^{(0)},g_{i}^{(n)}]
= [g_{j}^{(0)},g_{j}^{(n)}]  = \xi $ . From (50) we have $ (n+2) \xi = 0 $ .
Thus $ \xi = 0 $.

Now let us prove, that $ [g_{i}^{(l)},g_{i}^{(m)}] = 0 \; , \; l+m=n ,\; l\geq
1 $.

{}From  the Jacobi identity and the assumptions $ {\rm B}_{n} $  and  $ {\rm
C}_{n} $  we have:
\begin{eqnarray}
\sqrt{-1}[g_{i}^{(m)},g_{i}^{(n-m)}] = [g_{i}^{(m)}, [
g_{j}^{(0)},g_{k}^{(n-1-m)}]] = -\sqrt{-1} [g_{j}^{(0)},g_{j}^{(n)}]  -
\sqrt{-1} [g_{k}^{(n-1-m)},g_{k}^{(m+1)}] \; , \nonumber \\ 0 \leq m \leq n-1 .
\end{eqnarray}
hence $  [g_{i}^{(m)},g_{i}^{(n-m)}] =  [g_{k}^{(m+1)},g_{k}^{(n-m-1)}] , 0
\leq m \leq n-1 .$
Taking into account that, as has been proven, $  [g_{i}^{(0)},g_{i}^{(n)}]=0 $,
we obtain the statement of the lemma $ \Box $.

Thus it is shown, that the statements $ {\rm A}_{n+1},\; {\rm B}_{n+1}, \; {\rm
C}_{n+1} $  are true if the $ {\rm A}_{n},\; {\rm B}_{n}, \; {\rm C}_{n} $  are
true. From this the statements 2. and 3. a). of the theorem follow for the
subalgebra $ {\cal E}^{+} $ . For the subalgebra $ {\cal E}^{-} $ which is
isomorphic to $ {\cal E}^{+} $ a proof is identical.

By a direct calculation using the relations (22) one can verify that the
structure constants appearing in the RHS of the statement 3. b). satisfy the
recurrent relations (47),(37-39) with the initial conditions $ A_{k,0}^{(0,1)}
= \frac{1}{2}J_{ij} = w_{1}^{i}-w_{1}^{j}, \;  A_{k,0}^{(1,0)} =
-\frac{1}{2}J_{ij} $. This proves the statement 3. b)..

The statement 1. follows from the explicit form of the bases in $ {\cal E}^{+}
$ and $ {\cal E}^{-} $ .

The proof of the statements 3 c).,d). is very similar to that one of the
statements 3 a)., b)., for this reason  we describe it only   schematically.
Using induction one can derive recurrent relations for the structure constants
appearing in the decompositions of the commutators $ [g_{i}^{(l),\pm} ,
g_{i^{'}}^{(l),\mp} ] $ in the basis $ \{g_{i}^{(n),\pm} \} $ . Then, we
uniquely resolve these - the initial conditions being provided by the defining
relations (3-6). The relation (22) again plays the crucial role in this
solution.

To finish the proof of the theorem we need to check that the Lie bracket in $
{\cal E}_{k,\nu^{\pm}} $ as given by the formulas (11-14) satisfies the Jacobi
identity. This follows from the r-matrix representation (18,19) of the
commutation relations (11-14), and the well-known fact that $ r(u) $ (see (7))
satisfies the Classical Yang-Baxter equation $ \Box $.

\begin{large}
{\bf References} \end{large} \\
1. Reyman, A. G. and Semenov-Tyan-Shanskii, M. A., {\em Journ. Sov. Math.} {\bf
46}, 1631 (1989). \\
2. Belavin, A. A. and Drinfel'd V. G., {\em Funct. Anal. Appl. } {\bf 17}, 220
(1984).\\
3. Drinfel'd V. G., {\em Sov. Math. Doklady} {\bf 28}, 667 (1983).\\
4. Drinfel'd V. G., {\em Proceedings of the ICM}, Berkeley ,CA U.S.A., 798
(1986). \\
5. Chari, V. and Pressley, A., {\em L'Enseignement Math\'{e}matique} {\bf 36},
267 (1990); {Comm. Math. Phys.} {\bf 142}, 261 (1991). \\
6. Sklyanin, E. K., {\em Funct. Anal. Appl.} {\bf 16}, 263 (1983). \\
7. Kac, V. G., {\em Infinite dimensional Lie Algebras}, Cambridge University
Press, Cambridge, 1990. \\
8. Faddeev, L. D. and Takhtadjan, L. A., {\em Hamiltonian Methods in the theory
of solitons }, Springer-Verlag, Berlin, New-York, 1987. \\
9. Uglov, D. B., in preparation.

\end{document}